\documentstyle[mprocl]{article}

\def\be{\begin{equation}}
\def\ee{\end{equation}}
\def\bea{\begin{eqnarray}}
\def\eea{\end{eqnarray}}

\begin{document}

\title{METRIC FLUCTUATIONS IN SEMICLASSICAL 
GRAVITY$\mbox{}^{\hspace{0.1ex}\ast}$}

\footnotetext{$\mbox{}^{\ast\hspace{0.1ex}}$To appear in the 
proceedings of the 8th Marcel Grossmann Meeting (Jerusalem, Israel,
June 1997).}

\author{ R. MARTIN, E. VERDAGUER }

\address{Departament de F\'{\i}sica Fonamental, Universitat de
        Barcelona, Av. Diagonal 647,\\ \mbox{08028 Barcelona}, Spain}

\maketitle\abstracts{The Einstein-Langevin equation 
is a perturbative correction
to the semiclassical Einstein equation which takes into account 
the lowest order quantum fluctuations of the matter stress-energy
tensor. It predicts classical
stochastic fluctuations in the metric field which may 
describe some of the remnant gravitational fluctuations after the
process of environment induced decoherence driving the quantum to
classical transition of gravity.}


\section{Introduction}
\label{sec:introduction}


It is believed that there might be a regime in the early universe
in which gravity can be described classically, whereas the other
fields (``matter'' fields) must be treated quantically. 
One of the open problems in the theory of quantum fields 
is the study of the equations which describe the dynamics
of gravity in such ``semiclassical'' regime. 
The usual approach
to this problem is the description of the dynamics through the
semiclassical Einstein equation. However, some arguments have been often
given in the literature which make manifest that such description has
some limitations. 
One can argue that the classical metric appearing in the
semiclassical equation must correspond, from the point of view of a
quantum theory of gravity, to the expectation value in 
the semiclassical regime of some quantum metric operator. 
Thus, one can expect that, in general, this equation
will give some kind of mean value description of the 
classical spacetime geometry. This will be a
good description only in the limit where the fluctuations of the
spacetime geometry about its mean value are negligibly small. 
It is believed that this condition can break down when the state of
the matter fields is such that their stress-energy tensor has 
appreciable quantum fluctuations.\cite{ford,halliwell}
A better description in some of
these situations might be through the Einstein-Langevin equation,
which describe the dynamics of small stochastic corrections to the
solutions of the semiclassical Einstein equation.\cite{E-L,mv}


\section{The semiclassical Einstein equation}
\label{sec:semiclassical}


For simplicity, we shall consider matter as described by a linear
quantum field. 
Let $(M,g_{ab})$ be a globally hyperbolic spacetime and consider a
linear quantum field $\Phi$ on it, whose corresponding field operator
is $\hat{\Phi}[g]$. Let $|\psi\rangle$ be a physically acceptable
state (a state satisfying the Hadamard condition) of this field.
The set $(M,g_{ab},\hat{\Phi},|\psi\rangle )$ is a solution of the 
semiclassical Einstein equation if it satisfies
\be
{1\over 8 \pi G} \left( G_{ab}[g]+ \Lambda g_{ab} \right)-
2  \left( \alpha A_{ab}+\beta B_{ab} \right)\hspace{-0.3ex}
[g]=\left\langle
\hat{T}^{R}_{ab}[g]\right\rangle, 
\label{semiclassical Einstein eq}
\ee
where $\hat{T}^{R}_{ab}[g]$ is a renormalized stress-energy tensor
operator for the matter field quantized in the spacetime $(M,g_{ab})$ 
and the expectation value is taken in the state $|\psi\rangle$. 
In the above equation, $1/G$, $\Lambda /G$, $\alpha$ and $\beta$ are
renormalized coupling constants,
$G_{ab}$ is the Einstein tensor, and $A_{ab}$ and
$B_{ab}$ are the local curvature tensors obtained by varying with
respect to the metric the action terms corresponding to the Lagrangian
densities $C_{abcd}C^{abcd}$ and $R^2$, respectively
($C_{abcd}$ is the Weyl tensor and $R$ is the curvature scalar).
The stress-energy tensor for a given solution of the semiclassical
equation (\ref{semiclassical Einstein eq}) will in general have
quantum fluctuations. To lowest order, such fluctuations are described
by the bi-tensor, which shall be called noise kernel, defined as
\be
4 N_{abcd}[g](x,y) \equiv {1\over2}\, \biggl\langle  \biggl\{
 \hat{T}_{ab}[g](x)-
 \left\langle \hat{T}_{ab}[g](x) \right\rangle , \,
 \hat{T}_{cd}[g](y)- 
 \left\langle \hat{T}_{cd}[g](y)\right\rangle 
 \biggr\} \biggr\rangle,
\label{noise}
\ee
where $\{ \; , \: \}$ means the anticommutator, and where 
$\hat{T}_{ab}[g]$ denotes the unrenormalized stress-energy operator
(suitably regularized, with the regularization removed at the end of
the calculation). For a linear quantum field, this quantity is finite
and the operator $\hat{T}_{ab}$ can of course be replaced by the
renormalized one, $\hat{T}^{R}_{ab}$. If the value of the noise
kernel is appreciable, the description of the
gravitational field based on the semiclassical equation 
(\ref{semiclassical Einstein eq}) might break down. The quantum 
stress-energy fluctuations might induce classical stochastic
fluctuations in the metric field, which may be viewed as classical
remnants of the quantum fluctuations in the gravitational field 
after the process of environment induced decoherence driving the
quantum to classical transition of gravity.\cite{halliwell}


\section{The Einstein-Langevin equation}
\label{sec:E-L}


The Einstein-Langevin equation that we shall present in this section
give the first order perturbative correction to the semiclassical
Einstein equation (\ref{semiclassical Einstein eq}) which take into
account the lowest order stress-energy fluctuations, described by
(\ref{noise}).\cite{mv} 
We shall assume that such correction can
be described in terms of a linear perturbation $h_{ab}$ to the
semiclassical metric $g_{ab}$, solution of Eq.\ 
(\ref{semiclassical Einstein eq}). 
In order to give the dynamical equation for $h_{ab}$, let us
introduce a Gaussian stochastic tensor $\xi_{ab}[g]$
characterized by the following correlators:
\be
\left\langle\xi_{ab}(x) \right\rangle_{\xi} = 0,  
\hspace{10ex}
\left\langle\xi_{ab}(x)\xi_{cd}(y) \right\rangle_{\xi} = N_{abcd}(x,y),
\label{correlators}
\ee
where $\langle \hspace{1.5ex} \rangle_{\xi}$ means statistical 
average with respect to $\xi_{ab}$. The Einstein-Langevin equation for
the metric perturbation $h_{ab}$ is
\be
{1\over 8 \pi G} \Bigl( G_{ab}[g+h]+ 
\Lambda\left(g_{ab}+h_{ab}\right) \Bigr)- 
2 \left( \alpha A_{ab}+\beta B_{ab} \right)\hspace{-0.3ex}
[g+h]=\left\langle
\hat{T}^{R}_{ab}[g+h]\right\rangle+2 \xi_{ab}[g] , 
\label{Einstein-Langevin eq}
\ee 
where this last expression has to be understood as a correction to
(\ref{semiclassical Einstein eq}) linear in $h_{ab}$. The metric
perturbation $h_{ab}$, is a classical fluctuating
({\it i.e.}\ stochastic) field, that is, a field to which one assignes
a classical probability distribution functional. 
The mean value of the physical metric,
$g_{ab}+\left\langle h_{ab} \right\rangle_{\xi}$,
as we can see by taking the mean value of
Eq.\ (\ref{Einstein-Langevin eq}) with respect to $\xi_{ab}$, is a
solution of the semiclassical Einstein equation linearized
around $g_{ab}$. 
Notice that in Eq.\ (\ref{Einstein-Langevin eq}) the stochastic source
$\xi_{ab}[g]$ is not dynamical since it is defined through
the semiclassical metric $g_{ab}$ by the correlators
(\ref{correlators}). It is easy to show that it is covariantly
conserved up to first order in perturbation theory and it is thus 
consistent to include this tensor in
the right hand side of Eq.\ (\ref{Einstein-Langevin eq}). 
One can also see that for a conformal quantum field, {\it
i.e.}\ a field whose classical action is conformally invariant such as
a massless conformally coupled scalar field, one has that
$g^{ab}\xi_{ab}=0$. Thus, in this case, the trace of the right hand
side of Eq.\ (\ref{Einstein-Langevin eq}) comes only from the trace
anomaly. Being (\ref{Einstein-Langevin eq}) a stochastic equation for
$h_{ab}$, this metric perturbation can only be determined in a
probabilistic way. Thus we can only attempt to compute statistical
averaged quantities such as its correlation functions.


\section{Discussion}


We have introduced the Einstein-Langevin equation to give
a classical effective description of the effect of 
the lowest order stress-energy quantum fluctuations on the
spacetime geometry. A number of equations of the Langevin type for
some specific cosmological models have recently been given in the
literature using functional methods.\cite{E-L}
The two point correlation function of the metric 
perturbation have also been computed and discussed in some of these
works. The functional methods used
are actually motivated by the study of the process of
environment induced decoherence, which explains the appearence of 
effective classical equations
of motion for a field starting from the fundamental quantum theory of
this field in interaction with other quantum fields.\cite{halliwell}
One can show that
Eq.\ (\ref{Einstein-Langevin eq}) can be derived formally
using these methods.\cite{mv}


\section*{Acknowledgments}


This work has been
partially supported by the 
CICYT Project 
No.\ \mbox{AEN95-0882}, and the European Project 
No.\ \mbox{CI1-CT94-0004}.


\section*{References}



\begin{thebibliography}{99}

\bibitem{ford} 
  L.H. Ford, 
        {\it Ann.\ Phys.\ }{\bf 144}, 238 (1982);
  C.-I. Kuo and  L.H. Ford, 
        {\it Phys.\ Rev.\ }{\bf D47}, 4510 (1993),
  N.G. Phillips and B.-L. Hu, 
      {\sl Phys.\ Rev.\ }{\bf D55}, 6123 (1997).

\bibitem{halliwell} 
  J.J. Halliwell, quant-ph/9705005.

\bibitem{E-L}
  E. Calzetta and B.-L. Hu,
       {\sl Phys.\ Rev.\ }{\bf D49}, 6636 (1994);
  B.-L. Hu and A. Matacz,
       {\sl Phys.\ Rev.\ }{\bf D51}, 1577 (1995);
  B.-L. Hu and S. Sinha,
       {\sl Phys.\ Rev.\ }{\bf D51}, 1587 (1995);
  A. Campos and E. Verdaguer,
       {\sl Phys.\ Rev.\ }{\bf D53}, 1927 (1996);
  E. Calzetta, A. Campos and E. Verdaguer, 
       {\sl Phys.\ Rev.\ }{\bf D56}, 2163 (1997).

\bibitem{mv}
  R. Mart\'\i n and E. Verdaguer,
       in preparation (1997).

\end{thebibliography}
\end{document}